# Comparing EEG-Based Epilepsy Diagnosis Using Neural Networks and Wavelet Transform

**Mohammad Reza Yousefi** [1,*], **Saina Golnejad** [2], **Melika Mohammad Hosseini** [2], **Amin Dehghani** [3]

[1] IEEE Senior Member, Department of Electrical Engineering, Najafabad Branch, Islamic Azad University, Najafabad, Iran
[2] Digital Processing and Machine Vision Research Center, Najafabad Branch, Islamic Azad University, Najafabad, Iran
[3] Department of Psychological and Brain Sciences, Dartmouth College, Hanover, NH, United States
[*] Correspondence: mr-yousefi@iaun.ac.ir (Mohammad Reza Yousefi)

**Abstract:** Epilepsy is a common neurological disorder characterized by the recurrence of seizures, which can significantly impact the lives of patients. Electroencephalography (EEG) can provide important physiological information on human brain activity which can be useful to diagnose epilepsy. However, manual analysis and visual inspection of many EEG signals can be time-consuming and may lead to contradictory diagnoses by doctors. EEG signals play an important role in the diagnosis of epilepsy, as the quantification of cerebral signal anomalies may indicate the condition and the pathology of the cerebral signal. In this study, we attempted to develop a two-step process for the automated diagnosis of epilepsy using EEG signals. In the first step, we applied a low-pass filter and designed three intermediate filters for different frequency bands, and employed multi-layer neural networks. In the second step, we used a wavelet transform method to process the data. We also evaluated the use of two different classifiers, an artificial neural network (ANN) and a support vector machine (SVM), for the diagnosis of epilepsy. These classifiers were trained on normal and epilepsy data and were able to accurately distinguish between normal and epilepsy as well as other conditions. Our results showed that both the ANN and SVM had high accuracy, with 92.38% and 95.23% respectively. We also found that the use of the wavelet transform did not significantly affect the classification performance but using a multi-layer neural network provided better precision. In this study, we developed a two-step automated process, incorporating low-pass filters, intermediate filters, multi-layer neural networks, and wavelet transform, the researchers successfully achieved an accurate and efficient diagnosis of epilepsy. The results demonstrated high accuracy rates for both the artificial neural network (92.38%) and the support vector machine (95.23%) classifiers. Moreover, the study highlighted the effectiveness of utilizing a multi-layer neural network for improved precision. These findings contribute to the ongoing efforts in developing automated methods for epilepsy diagnosis, offering the potential for faster and more reliable detection techniques that can enhance patient care and outcomes.

## 1. Introduction

Epilepsy, a neurological disorder impacting approximately 50 million people worldwide, witnesses the addition of 2.4 million new cases each year. Developing countries bear a significant burden, with two-thirds of new patients hailing from these regions. As the global population ages and the prevalence of mental and neurological disorders rises, epilepsy and similar conditions are anticipated to emerge as major public health concerns. Consequently, the development of tools, such as computer-assisted diagnosis systems, becomes crucial in supporting neurologists and psychiatrists in accurate epilepsy diagnosis. Furthermore, epilepsy can be classified based on various factors, including etiology, type of seizure, or syndrome. The World Health Organization reports that 20% of patients have primary epilepsy, while the remaining 80% experience focal epilepsy (Liu et al., 2021). Epilepsy, characterized by brain seizures, can be categorized as either focal or non-focal. Focal seizures originate from abnormal brain cells, whereas non-focal seizures arise from normal brain cells. To detect epilepsy accurately, specialized signal processing techniques have been devised to differentiate between focal and non-focal seizures. Additionally, careful analysis of focal signals is crucial, particularly for patients being considered for surgery. This ensures comprehensive evaluation and informed decision-making regarding surgical intervention. The severity of epilepsy can be classified as "early" or "progressive", with patients in the latter category requiring surgery. Decomposing the electroencephalogram (EEG) signal can help clinicians and neurophysiologists to identify focal signals by extracting characteristics from each sub-band layer (Rajagopalan & Rajagopal, 2020). Automatic identification and detection of epilepsy or brain-related disorders is a major concern for doctors and researchers. Neurologists and brain surgeons often rely on the interpretation of EEG signals to diagnose epilepsy. Therefore, it is essential to develop a reliable and automated technique for detecting epilepsy attacks. The automatic interpretation of EEG signals in the diagnosis and treatment of brain diseases is one of the most important areas of research. There are various neuroimaging techniques for measuring brain activity, such as electroencephalography (EEG), magnetoencephalography (MEG), functional magnetic resonance imaging (fMRI), and functional near-infrared spectroscopy (fNIRS). EEG is commonly used to capture brain activity through sensors called electrodes, as neurons communicate through electrical signals that reach the surface of the brain (Chen et al., 2020; Dehghani et al., 2020). Researchers from various interdisciplinary fields, including engineering, neurology, microelectronics, bioengineering, and neurophysiology, are using EEG-based signal processing technology for a variety of applications, such as controlling external devices, communication, and medical diagnosis. This technology is particularly useful for diagnosing and monitoring neurological brain disorders because it allows for the non-invasive detection of electrical activity in the brain. One common application of EEG is in the diagnosis of epilepsy (Alturki et al., 2021).

Epilepsy is defined as a condition characterized by recurrent seizures that sometimes caused by an external stimuli. According to research, about 4% of people worldwide will experience epilepsy seizure at some point in their lives, with 1% developing the condition. Epilepsy is associated with abnormal brain activity in the central nervous system (CNS) (Kocadagli & Langari, 2017). Epilepsy is a neurological disorder characterized by abnormal brain activity that can occur in one part or throughout the entire brain. The result of this irregular activity is seizures, loss of consciousness, exaggerated emotions, and abnormal behavior. Epilepsy can affect anyone, regardless of gender, age, or ethnicity. The main symptoms of epilepsy include staring, confusion, uncontrollable shaking, loss of consciousness, and psychological symptoms such as fear (Drenthen et al., 2021).

Epileptic seizures are classified based on the part of the brain where abnormal brain activity begins. According to the International League Against Epilepsy (ILAE), a diagnosis of epilepsy requires at least one seizure without a known trigger, or a risk of recurrent seizures or an epilepsy syndrome. Pediatric epilepsy can be particularly complex due to the diverse expressions of syndromes that require diagnosis, evaluation, and specific treatment. Children with epilepsy may differ from adults with the condition, particularly in the expression of age-specific epileptic syndromes. This requires careful consideration of factors such as diagnosis, classification of epilepsy, evaluation to determine the pathology, and decision-making for treatment. A thorough understanding of these factors can help researchers make an accurate diagnosis of epilepsy and guide appropriate testing and treatment decisions (Kocadagli & Langari, 2017).

Epilepsy is a serious disorder characterized by recurrent seizures due to sudden dysfunction in the brain. These seizures can disrupt normal brain function and may cause other side effects such as amnesia, depression, and other mental disorders. Early identification of epilepsy is important in order to take appropriate action to prevent potential consequences and ensure the patient's health. In recent decades, there has been active research on the automatic detection of seizures and the use of EEG signals for the detection of epilepsy (Ebrahimzadeh, Shams, et al., 2019; Ebrahimzadeh, Soltanian-Zadeh, et al., 2019). EEG is a non-invasive method commonly used in the clinical diagnosis of epilepsy. However, it can be tedious and time-consuming for neurologists to manually identify seizures from EEG recordings. As a result, there is a need for the development of a reliable automatic epilepsy detection system that can significantly improve the quality of treatment for patients with epilepsy (M. Sharma et al., 2020; Tuncer et al., 2021).

Epilepsy encompasses various categories determined by the nature of symptoms and the level of brain involvement, including etiology, type of seizure, and syndrome. Generalized epilepsy is characterized by epileptic waves that span the entire brain and cause widespread effects, whereas focal epilepsy is associated with epileptic waves that impact a specific region of the brain. However, it is important to note that the term "epileptic waves" is not well-defined and requires further clarification (Ullah et al., 2018). In this study, the performance of epilepsy predictors will be evaluated in terms of metrics such as sensitivity, specificity, and accuracy using two methods: one that employs wavelet transform and one that does not. A system will then be designed that can detect epilepsy by receiving EEG signals. The predictors will be evaluated separately using both methods.

This study will follow these steps: first, a review of previous studies will be provided. Then, the datasets used in the study will be described. The proposed method, including preprocessing, signal processing, feature extraction, and classification, will be explained. In conclusion, a comparative analysis will be conducted to evaluate the effectiveness of the proposed method in relation to other approaches. The proposed method employed two distinct classifiers, Multilayer Perceptron (MLP) and Support Vector Machine (SVM), to accurately classify and differentiate between healthy subjects and patients. The obtained results will shed light on the performance and potential advantages of this approach compared to alternative methods.

## 2. Related Studies

This article highlights several noteworthy studies in the field of epileptic seizures, focusing on their significance and relevance. (Tzimourta et al., 2018) et al. have proposed a method for automatic seizure detection in EEG recordings using the discrete wavelet transform (DWT) and support vector machines (SVM). The method consists of four steps: segmentation, wavelet analysis, feature extraction, and classification. In the first step, the long-term EEG recording is divided into 2-second windows. Then, 5-level wavelet analysis is applied to each window, dividing the signal into several frequency sub-bands. In the next step, five features are extracted from each sub-band and used to train an SVM classifier. The classifier is then used to classify the sub-bands as either seizure or non-seizure activity. In the study by (Jaiswal & Banka, 2018), two techniques using SVM were employed to classify EEG signals as convulsive or non-convulsive. The first technique, called sub-pattern-based principal component analysis (SpPCA), involves dividing the input patterns into subsets and extracting features from each subset using principal component analysis (PCA). The extracted features are then combined according to the pattern partition sequence to form the final feature vectors. The second technique involves performing PCA on the features extracted in the first step to further reduce the dimensionality and extract overall features. In the study by (Shoeb, n.d.) , they used SVM to diagnose epileptic seizures using EEG data. Their approach achieved an accuracy of 96% when tested on experimental data.

(Rahul Sharma et al., 2019) used a third-order cumulant function to automatically detect focal EEG signals. They extracted features from the EEG signals using locality sensitive discriminant analysis (LSDA) and then SVM to classify these features. The authors obtained a maximum classification accuracy of 99% using the Bern-Barcelona EEG dataset.

(Siddharth et al., 2019) proposed a method for distinguishing between focal and non-focal EEG signals using sliding mode singular spectrum analysis. The authors calculated features from the reconstructed component of the EEG signal and used a radial basis function neural network to classify them. When tested on the Barcelona EEG dataset, the method achieved an average accuracy of 99.11%, average sensitivity of 98.52%, and average specificity of 99.7%.

(Rajeev Sharma et al., 2014) proposed a method to extract 25 features of the EEG signal using the LSDA method. These features were then classified using an SVM classification approach. The authors tested the method using the cross-validation approach on the Barcelona EEG dataset, achieving a classification accuracy of 96.2%.

(Nigam & Graupe, 2004) proposed an EEG-based computer diagnostic method for epilepsy using a combination of nonlinear filters and an artificial neuron network (ANN). The proposed method achieved an accuracy of 97.2%.

(Kannathal et al., 2005) compared various entropy algorithms and found that entropy values could be used to distinguish between normal and epileptic EEG. They used an adaptive neuro-fuzzy inference system (ANFIS) for classification and achieved 92.2% precision.

(Sadati et al., 2006) used an adaptive neural fuzzy network and the energy of discrete wavelet transform (DWT) sub-bands for epilepsy diagnosis. However, their proposed method had a low accuracy of approximately 85.9%.

(Dalal et al., 2019) studied the use of the flexible analytic wavelet transform (FAWT) to decompose EEG signals. This non-stationary transform generated fractal dimension features at each scale level. The proposed method was tested using the Kruskal-Wallis statistical test, and the authors achieved an average classification accuracy of 89.1% on the Barcelona EEG dataset.

(Deivasigamani et al., 2016) analyzed a soft computing-based adaptive neuro-fuzzy inference system (ANFIS) classification method and used the feature values to distinguish between focal and non-focal signals. Out of 700 EEG signals, their method correctly classified 694 signals, resulting in a detection rate of 99.1%. (Abhinaya & Charanya, 2016) developed a method to extract features based on the entropy of the EEG signal and optimized them using a selection method. They applied a linear regression model with an SVM classifier to the optimized features to distinguish between focal and non-focal signals and achieved a classification rate of 92.8% using an open-access EEG dataset. (Ibrahim et al., 2018) used a wavelet and Shannon entropy-based method to recognize epilepsy seizures and applied SVM, LDA, artificial neural networks, and k-nearest neighbor as classifiers for their proposed system.

(Gruszczyńska et al., 2019) proposed a study using recurrence quantification analysis to classify EEG signals. The results of the study were visualized using principal component analysis and classified using an SVM classifier. Anuragi et al., developed a machine learning-based approach using wavelet transform to classify individuals with alcohol use disorder. The study used SVM and Naïve Bayes methods (Anuragi et al., 2021). (Mutlu, 2018), proposed a method using the decomposition of Hilbert-Huang epileptic vibration and used the minimum SVM classifier to distinguish between normal and epileptic EEG signals. (Al-Salman et al., 2019) used a wavelet Fourier analysis to detect sleep signals in EEG and applied a least-square SVM classifier. (Zarei & Asl, 2021) used discrete wavelet transform (DWT) and orthogonal matching pursuit (OMP) techniques to extract coefficients from EEG signals, then calculated nonlinear features and statistical features using these coefficients. They evaluated the performance of the proposed techniques using three commonly used EEG datasets. (Yavuz et al., 2018), proposed the use of Mel frequency cepstral coefficients (MFCCs) and generalized regression neural networks to distinguish normal and seizure EEG recordings. (Omidvar et al., 2021) used discrete wavelet transform (DWT) to divide EEG signals into sub-bands and extract features. They also used genetic algorithms to select effective features and applied artificial neural network (ANN) and vector support machine (SVM) classifiers to the data. The simulation results showed that the proposed method had higher accuracy in detecting epilepsy attacks compared to other similar approaches.

**3. Methods and Materials**

*3.1. Data and Materials*

To test the method in this study, epilepsy data from the Bonn University database was used. This database includes 5 different recording models, each with 500 pieces of 100 points: (A) non-epileptic recording with open eyes, (B) non-epileptic recording with closed eyes, (C) includes recordings of EEG from the hippocampal formation in the hemisphere opposite the epileptogenic zone and (D) comprise EEG recordings of the epileptogenic zone and finally (E) is a collection of epileptic seizure activity recorded from the hippocampal focus (Tanke et al., 2019). The length of each recording is 23.6 seconds and the sampling frequency is 173.6 Hz. In this study, models B and D were used, both with and without wavelet transform decomposition at a level of 8. The details of the database are summarized in Table 1. The database is open-access and available on the website of Bonn University.

Table 1. The details of the database

| Set | Patients |
|---|---|
| A | Healthy |
| B | Healthy |
| C | Epilepsy |
| D | Epilepsy |
| E | Epilepsy |

*3.2. Proposed Method*

After obtaining the data from the mentioned database, data preprocessing was started. A 70 Hz low-pass filter was applied to both methods. Then, the alpha, beta, and gamma frequency bands were extracted using separate mid-pass filters. In the feature extraction step, the Frequency median (FMD), frequency mean (FMN), frequency ratio (FR), and Waveform length (WL) features were extracted (Dehghani et al., 2021, 2022a, 2022b; Mosayebi et al., 2022; Yousefi, 2022). Finally, a multilayer perceptron neural network (MLP) classifier was designed and applied to the data. After obtaining satisfactory results with this classifier, the wavelet transform method was applied separately. The diagram of the proposed method is shown in Fig. 1.

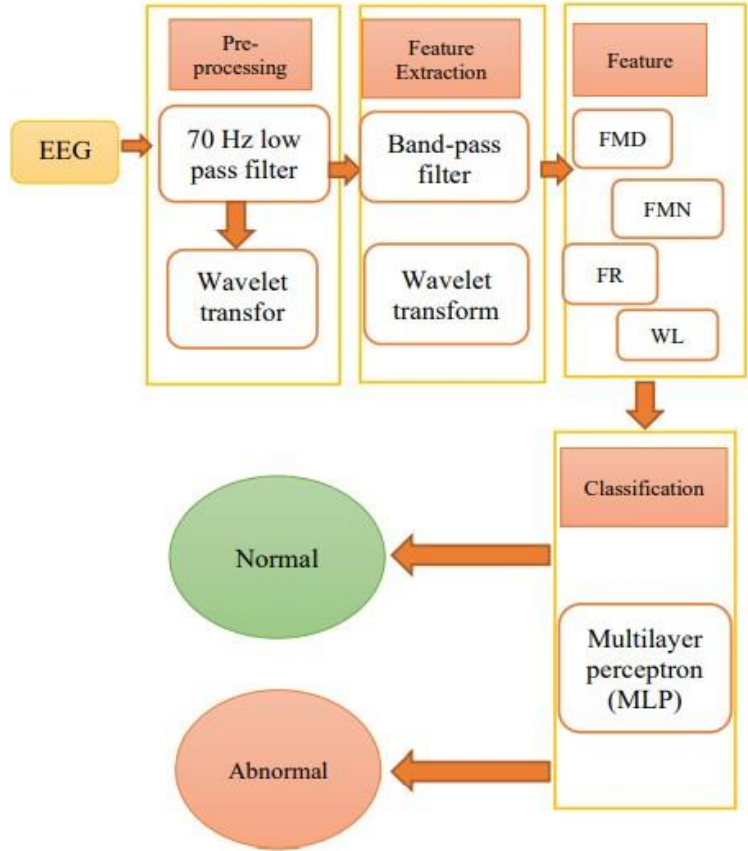

Fig. 1. Diagram of the proposed method

*3.3. Wavelet Transform*

The wavelet transform is a mathematical transformation widely used in various scientific fields, and can be represented by the following general equation [1]:

$$f(x) = \frac{a_0}{2} + \sum_{n=0}^{\infty}(a_n \cos w_n + b_n \sin w_n) \qquad (1)$$

The properties of a wavelet function (Ψx) include:

(A) being appropriately limited in time

(B) having a mean of zero. $\int_{-\infty}^{+\infty} \Psi t \, dt = 0$

C) has a non-zero state: $0 < \int_{-\infty}^{+\infty} |\Psi x|^2 \, dt < \infty$

*3.4. Signal Preprocessing*

In the preprocessing step, a 70 Hz low-pass filter was applied to the raw EEG signal collected in both stages, with and without the use of wavelet transform. The FDATOOL toolbox was used to design the filter, and the wave menu tool in the MATLAB 2020 software was used for wavelet transform. To remove noise using wavelet transform, the de-noise section of the wavelet transform toolbox was used. This section allows the wavelet to independently identify noise by comparing the daughter signal (EEG signal) to the mother signal (wavelet signal). If they do not match, the details from the lower levels are removed while preserving the generalities.

*3.5. Signal Processing*

In this step, a band-pass filter with Fpass1 = 25 Hz, Fstop2 = 71 Hz, Fstop1 = 24 Hz was applied to the alpha, gamma, and beta bands, and a separate band-pass filter with minimum order values of Fpass2 = 70 Hz and Fstop = 71 Hz was applied to the gamma band. The passband and stopband ripple were set to 0.1 and 0.01, respectively. To process the signal using wavelet transform, the filter coefficients, including approximation and detail, were obtained and the detail and general coefficients at different levels were calculated separately for the alpha-beta and gamma bands. The output signal for these coefficients was calculated using the 'wrcoef' command in the time domain in MATLAB.

*3.6. Feature Extraction*

Feature extraction is an essential step for detecting epilepsy attacks. It is used to establish epilepsy data using standard and collected epilepsy data. The wavelet transform is often used to decompose a signal into scaled and translated versions of a mother wavelet and a scaling function. The discrete wavelet transform (DWT) has been frequently used in epileptic seizure detection with promising results. Enhancing the precision of epilepsy detection heavily relies on improving the accuracy of feature extraction. The power spectrum, which represents the relative magnitude of frequency components comprising a signal, plays a vital role in this process. It is essential that the data utilized to determine the power spectrum adequately captures the signal's excitation levels to ensure reliable results. In this study, four important frequency and time features were extracted separately for epileptic and non-epileptic data in the alpha, gamma, and beta frequency bands: FMD, FMN, FR, and WL. These features were extracted from EEG recordings of stimulation with open and closed eyes. The extracted features are listed in Table 2.

Table.2 Extracted feature

| Number | Frequency Domain Features | Number | Time Domain Features |
|---|---|---|---|
| 1. | Frequency median (FMD) | 1 | Waveform length (WL) |
| 2. | Frequency mean (FMN) | 2 | - |
| 3. | Frequency rate (FR) | 3 | - |

In individuals with epilepsy, the median and mean frequency increase due to higher brain frequency, and the frequency ratio decreases. The Waveform length feature measures changes and calculates the absolute value of the next sample minus the previous sample. If the signal is smooth with little change, this property becomes zero. Therefore, in epilepsy, where the disorder entropy is low and brain oscillations are consistent, the Waveform length decreases. Shannon entropy is used to calculate the entropy of wavelet transform, which measures the entropy of changes.

*3.7. Classification*

Two different classifiers were used in this study: the Multilayer Perceptron (MLP) and Support Vector Machine (SVM), to classify and differentiate between healthy subjects and patients. A ten-fold cross-validation method was used to evaluate the performance of the classifiers. The EEG signals were classified into ten parts with equal numbers of signals, except for two groups. One part was used for testing and the other nine parts were used for training the classifier. This process was repeated ten times for each different test part, and the average performance for accuracy, sensitivity, and specificity was calculated. Each of the ten processes was repeated fifteen times to improve the accuracy.

*3.7.1 Multilayer Perceptron (MLP)*

The Multilayer Perceptron neural network was trained using the backpropagation algorithm and a variable learning rate to diagnose the disease. The input layer had the same number of nodes as the length of the input vector for each time interval, and the output layer had one node, which represented the possibility of only two classes being classified. The number of neurons in the hidden layer was selected to achieve an optimal architecture. It is important to note that the training process was only performed on the training data to ensure that the network did not observe the testing data when selecting the optimal structure, which allows for effective generalization of the network. The network was tested using the testing data once the training error reached a minimum. A linear transfer and a sigmoid function were used as the output node and the hidden layer, respectively. The network training continued until the mean value was much less than 0.01 or the range of training iterations reached 1000.

*3.7.2. Support Vector Machine*

Support vector machine (SVM) is a popular supervised learning method that maps data from an N-dimensional input space to an M-dimensional feature space (M > N) in order to linearly separate classes. Based on statistical learning theory, SVM is considered an extension of the generalized portrait algorithm, which analyzes data for classification and regression. In this article, the Support Vector Machine (SVM) classifier was employed due to its widespread usage in data classification, pattern recognition, and regression analysis. SVM is a supervised classifier known for its effectiveness in these domains. The purpose of regression is to determine the best version out of a set of models, known as estimating functions, to accurately approximate target values. The standard support vector regression estimating function is:

$$f(x) = (w \cdot \Phi(x)) + b \quad (2)$$

where $w \subset R_n$, $b \subset R$ and $\Phi$ are a nonlinear function that maps x into a higher dimensional space. W and b are the weight vector and bias, respectively. The weight vector (w) can be expressed as:

$$w = \sum_{i=1}^{L}(\alpha_i - \alpha_i^*) \quad (3)$$

By substituting Eq. (2) into Eq. (3), the generic equation can be rewritten as:

$$f(x) = \sum_{i=1}^{L}(\alpha_i - \alpha_i^*)(\Phi(xi) \cdot \Phi(x)) + b \quad (4)$$

$$f(x) = \sum_{i=1}^{l}(\alpha_i - \alpha_i^*) k (xi.x) + b \quad (5)$$

In Eq. (5), the function k ($x_i$ . x) = ($\Phi(x_i)$. $\Phi(x)$) is replaced with the dot product and is referred to as the kernel function and $\alpha = (\alpha\ 1, \alpha\ 2, …, \alpha\ l)$ consists of nonnegative Lagrange multipliers. The selection of kernel functions and kernel parameters depends largely on the specific application. Some commonly used kernel functions include radial basis functions (RBFs) and polynomial kernel functions. The formulas for these kernel functions are shown below, respectively:

$$\frac{-|x-xi|^2}{2\sigma^2} \tag{6}$$

$$[(x * x_i) + 1] \tag{7}$$

where, "σ" represents the kernel width and "d" represents the order, which were experimentally determined to produce the best classification results. In this work, RBFs and polynomial kernel functions were used with various values of sigma (σ = 0.8, 1, 1.2) and orders (d = 1, 2, 3), respectively.

*3.8. Evaluation*

To assess the effectiveness of the proposed method for diagnosing epilepsy, we use common performance measures such as accuracy (AC), sensitivity (SN), specificity (SP), and precision (P). In Equations (25) to (28) below, TP refers to true positives (correctly diagnosed epilepsy), TN refers to true negatives (correctly diagnosed non-epilepsy), FN refers to false negatives (incorrectly diagnosed as non-epilepsy), and FP refers to false positives (incorrectly diagnosed as epilepsy).

$$AC = \frac{TP+TN}{TP+TN+FN+FP} \tag{8}$$

$$SN = \frac{TP}{FN+TP} \tag{9}$$

$$SP = \frac{TN}{TN+FP} \tag{10}$$

$$P = \frac{TP}{FP+TP} \tag{11}$$

**4. Results**

The results of the validation of the 10-layer neural network used to classify the signal with 95.5% accuracy are shown in Fig. 2. Next, wavelet transform with level 8 and decomposition level 4, along with Shannon entropy, were used to diagnose epilepsy with 91% accuracy. The validation results are shown in Fig. 3. We applied the proposed method to differentiate patients from normal subjects by building two individual classifiers using features extracted from the EEG signals. The first classifier is a Support Vector Machine (SVM) implemented with a linear kernel, which achieved 95.23% accuracy, 98.78% specificity, 82.60% sensitivity, and 95% precision. The second classifier is a Multilayer Perceptron (MLP) classifier, which achieved 91.90% accuracy when trained on 90% of the EEG data and tested on the remaining 10%. The statistical measures for validation by the SVM and MLP classifiers are shown in Table 2, and the results for each classifier are summarized in Table 3. These results demonstrate the effectiveness of the proposed method in classifying the two classes using the mentioned classifiers. Table 4 shows the results of the performed method which confirms thee results of Table 3.

Table 5 presents a comprehensive comparison of various studies to validate and reinforce the proposed method. It serves to support the stability and reliability of the proposed approach by highlighting its performance in relation to other relevant studies.

To optimize the learning cost and diagnosis performance, it is important to carefully choose the SVM classifier parameters and kernel width. To do this, we divided the training data into train and validation sets and selected the optimum values of the parameters such that the smallest error on the validation dataset is achieved. The optimum value of the parameter delta was found to be 1.5.

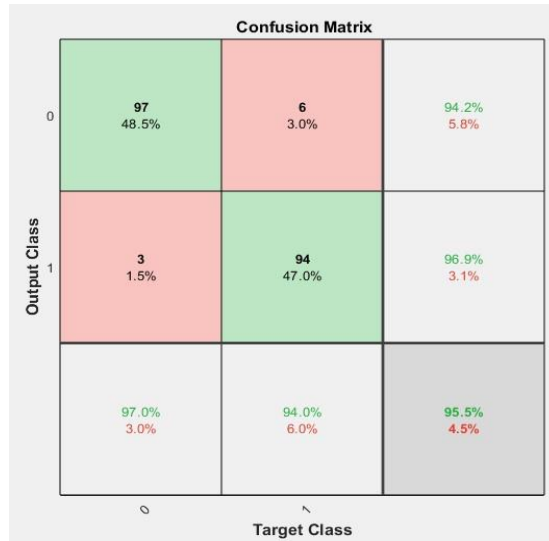

Fig. 2. The validation value and accuracy of perceptron multilayer neural network

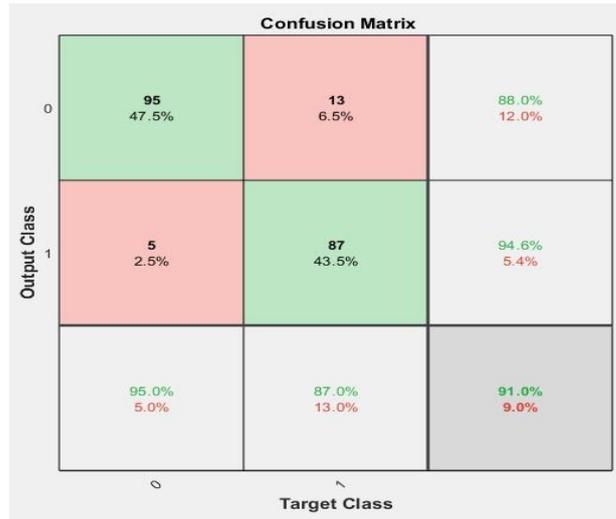

Fig. 3. The validation value and accuracy after wavelet transform conversion Table.1 Overview of Bonn dataset

Table.3 Results of the classifier

| Methods | Accuracy | Sensitivity | Specificity |
|---|---|---|---|
| Wavelet Transform | 91.0% | 95% | 87% |
| Multilayer perceptron neural network | 95.5% | 97% | 94% |

Table.4 Results of the performed method

| Methods | Accuracy | Sensitivity | Specificity |
|---|---|---|---|
| wavelet transform | 91.0% | 95% | 87% |
| Multilayer perceptron neural network | 95.5% | 97% | 94% |

Table.5 Comparison between different methods

| Methods | Accuracy |
|---|---|
| Proposed Method using Wavelet Transform | 91.0% |
| Proposed Method using Multilayer perceptron neural network | 95.5% |
| adaptive neuro-fuzzy inference system (ANFIS) (Kannathal et al., 2005) | 92.2% |
| flexible analytic wavelet transform (FAWT) (Dalal et al., 2019) | 89.1% |
| adaptive neural fuzzy network and the energy of discrete wavelet transform (DWT) (Sadati et al., 2006) | 85.9% |

## 5. Discussion

As previously mentioned, the causes of epileptic seizures include epidemiology, etiology, acute stroke, brain infection, and the adverse effects of certain medications. In past studies, some researchers have used a combination of EEG signals with FMRI to detect physiological values (Ebrahimzadeh, Shams, et al., 2019; Ebrahimzadeh, Soltanian-Zadeh, et al., 2019). In recent years, advanced methods focusing on functional dynamics have been proposed in studies to assist people with epilepsy. For example, the SL parameter in the theta frequency band has been studied in various studies of individuals with epilepsy, and has been shown to classify epilepsy with a sensitivity of 62% and a specificity of 76%. The techniques for analyzing brain signals have been improved in recent years as they can reflect brain activity. The aim of this study was to demonstrate the predictive value of physiological signals measured using EEG, which was conducted in two different experimental stages. By testing on the Bonn datasets, this study also showed competitive results compared to other recent methods in terms of how a predictive model based on EEG characteristics can significantly improve upon clinical diagnosis and aid in the diagnosis of epileptic symptoms in the EEG signal. In the first step, using a multilayer perceptron neural network (MLP) classifier, epilepsy and control signals in the alpha, beta, and gamma frequency bands were analyzed separately and 95% accuracy was achieved in detecting the epileptic signal. Then, using a wavelet transform function, the signal in the alpha, beta, and gamma frequency bands was re-examined and approximately 91% accuracy was achieved in detecting the epileptic signal. This study has succeeded in developing a simpler method for assessing epilepsy from EEG signals and also achieved satisfactory parameters such as accuracy, compared to other studies.

## 6. Conclusion

Visual examining long-term EEG recordings to detect epileptic seizures is a costly and time-consuming method, so this study proposed a new seizure detection algorithm using EEG signals to address these issues. The main strength of this study is the analysis of the available sample using both a multilayer perceptron neural network classifier and a wavelet transform function.

Future studies with larger sample sizes are needed to confirm and expand upon these findings. In this study, the diagnostic value of physiological signals measured using EEG was evaluated, and features such as FMD, FMN, FR, and WL were extracted. FMD and FMN, which are the median and mean frequencies, respectively, are obtained through the Fourier transform of the desired signal. In individuals with epilepsy, these two characteristics increase due to higher brain frequency compared to normal individuals. FR, or frequency rate, refers to the high-to-low frequency ratio of the brain, and in individuals with epilepsy, the frequency rate decreases due to higher brain frequency compared to normal individuals. WL, or waveform length, is a temporal feature that measures the changes and complexity of the signal by subtracting the absolute magnitude of subsequent samples from the previous samples and adding them together. In individuals with epilepsy, WL is reduced due to decreased entropy and similar fluctuations in the brain. Calculating these features provides useful information for physicians to diagnose and treat patients with epilepsy.

According to the studies conducted in this paper, although there is not a significant difference between the two methods used, the experiment using the multilayer perceptron neural network (MLP) achieved slightly higher accuracy than the experiment using the wavelet transform function, as shown in Table 3. In recent years, new methods for automatic epilepsy detection have been emerging. Improving the speed and accuracy of epilepsy detection models will contribute to the development of clinical diagnosis techniques and portable, integrated epilepsy detection equipment. Therefore, a concise and efficient epilepsy detection model will be a necessary trend in the future.